\documentclass[reprint
, amssymb, amsmath
,aps, prb
, floatfix
,superscriptaddress]{revtex4-1}

\usepackage{bm}
\usepackage{graphicx}
\usepackage{dcolumn}
\usepackage{titlesec}
\usepackage[colorlinks=true, linkcolor=blue, urlcolor=blue, citecolor=blue]{hyperref}

\begin{document}

\titlespacing{\section} {0pt}{3ex plus 1ex minus 1ex}{2ex plus .5ex}
\titlespacing{\subsection} {0pt}{3ex plus 1ex minus 1ex}{2ex plus .5ex}

\title{Universal Correlation between Critical Temperature of Superconductivity and band structure features}
\author{Yang Liu}
\email{yliu@iphy.ac.cn}
\email{l\_young@live.cn}
\affiliation{Institute of Physics, Chinese Academy of Sciences, Beijing, 100190, P.R. China}
\affiliation{School of Materials Science and Engineering, University of Science and Technology Beijing, Beijing, 100083, P.R. China}
\author{Ning Chen}
\email{nchen@sina.com}
\affiliation{School of Materials Science and Engineering, University of Science and Technology Beijing, Beijing, 100083, P.R. China}
\author{Yang Li}
\email{yang.li@upr.edu}
\affiliation{Department of Engineering Science and Materials, University of Puerto Rico, Mayaguez, Puerto Rico 00681-9000, USA}

\date{\today}
\begin{abstract}
The critical temperature (${T}_\text{c}$) of superconductors varies a lot. The factors governing the ${T}_\text{c}$ may hold key clues to understand the nature of the superconductivity. Thereby, ${T}_\text{c}$-involved correlations, such as Matthias laws, Uemura law, and cuprates doping phase diagrams, have been of great concern. However, the electronic interaction being responsible for the carriers pairing in high-${T}_\text{c}$ superconductors is still not clear, which calls for more comprehensive analyses of the experimental data in history. In this work, we propose a novel perspective for searching material gene parameters and ${T}_\text{c}$-involved correlations. By exploring holistic band structure features of diverse superconductors, we found a universal correlation between the ${T}_\text{c}$ maxima and the electron energy levels for all kinds of superconducting materials. It suggests that the ${T}_\text{c}$ maxima are determined by the energy level of secondary-outer orbitals, rather than the band structure nearby the Fermi level. The energy level of secondary-outer orbitals is a parameter corresponding to the ratio of atomic orbital hybridization, implying that the fluctuation of the orbital hybridization is another candidate of pairing glue. 
\end{abstract}
\maketitle

\section{Introduction}

Superconductivity is such an intriguing phenomenon that its microscopic mechanism remains mysterious. 
Especially, there is even a lack of consensus on what and how many unvoidable problems need to be solved to understand the mechanism of high temperature superconductivity (HTS)\cite{473}.

So far, the pairing mechanism in the HTS is not identified. 
Whether the pairing of carriers is supported by a kind of boson quasi-particle (pairing glue), or by the direct interaction between the carriers themselves? 
Does it need a boson glue, what is the glue (if exists), and how can we confirm it? The debate is still going on\cite{38}. 
Although the experimental works have provided many clues to the pairing glue or the pairing interaction, there is no conclusion with solid evidences, yet. 

Anyhow, one big challenge is to interpret how the cuprates superconduct, and another is why their superconducting critical temperature (${T}_\text{c}$) are so high.
In recent years, a lot of new superconducting materials have been discovered, but the ${T}_\text{c}$ record at ambient pressure is still held by cuprate. 
It leads to a series of questions:
What makes cuprates so special?
What is (are) the vital factor(s) governing the ${T}_\text{c}$? 
And what causes the elevated ${T}_\text{c}$ in cuprates? 

Nowadays, it is of growing significance to study the variation of ${T}_\text{c}$ with various parameters. 
The well-known phase diagrams\cite{429, 457, 198} of the high-${T}_\text{c}$ superconductors have indicated that doping can greatly affect the ${T}_\text{c}$. 
It is a common characteristic that for each high-${T}_\text{c}$ superconducting material, the ${T}_\text{c}$ varies with doping, and reaches a maximum at the optimal doping content (\textit{i.e.} the optimal  ${T}_\text{c}$). 
However, the doping content is not a physical parameter. 
Several true physical parameters (\textit{e.g.} carrier concentration, strength of the interaction between carriers) are hidden behind, varying with the doping content all together. 
So far, it is hard to obtain the specific values of those physical parameters in diverse materials, both experimentally and theoretically.
Then, the variation of ${T}_\text{c}$ with each of those physical parameters is not so clear, yet.
Besides, the whole phase diagrams are intricate. 
Various anomalies in the normal state, as well as the relationship between the superconducting phase and  other phases with different orders, need to be explained.
Therefore, it took years, and will take much more, to understand what the phase diagram tells.

Meanwhile, the optimal ${T}_\text{c}$ of diverse high-${T}_\text{c}$ superconducting materials are quite different, which means some factor other than doping is vital. 
In another word, the factors governing the ${T}_\text{c}$ and the factors governing the optimal ${T}_\text{c}$ are not the same. 
Therefore, fortunately, we can focus on the material diversity of the optimal ${T}_\text{c}$, which also can provide important information on the HTS mechanism.

Moreover, given the complexity of the HTS problem and the diversity of the superconducting materials, the smoking gun evidence for the HTS mechanism seems not coming from several isolated experiments on a few archetypal materials, but should be derived from a statistical result of massive experiments on superconducting materials as many as possible. 
It needs a holistic perspective, considering not only the diversity of high-${T}_\text{c}$ superconductors, but also the diversity of all superconductors. 
From this perspective, we need universal laws to indicate how the pairing interaction varies in diverse superconducting materials, and how it affects the ${T}_\text{c}$. 
Also, the exploration of new superconductors is counting on it.

A good universal law is both a clue leading to the HTS mechanism and a guidance for predicting new superconducting materials. 
In this work, we first review a few landmark ${T}_\text{c}$-involved laws, and make a discussion on how and where to find universal laws with material gene parameters. 
Then we present a universal law showing the relevance between the ${T}_\text{c}$ and a band structure parameter, which gives an unexpected clue to the pairing mechanism in the HTS.

\section{A brief review on ${T}_\text{c}$-involved laws of superconducting materials}

Ever since the superconductivity was discovered, plenty of ${T}_\text{c}$-involved laws or correlations have been found\cite{196, 215}. 
Some of them are known as empirical material laws, which do not care about the physical picture, but faithfully show the dependence of superconducting ${T}_\text{c}$ on material parameters. 
For example, three Cu-O planes per unit cell being better for cuprates\cite{215}, the ${T}_\text{c}$ varying with the bond lengths and angles within the Cu-O planes or Fe-As(Se) layers\cite{139, 199, 311, 69, 214}, as well as Matthias laws for the conventional superconductors\cite{337}. 
Besides, the laws found with the aid of machine learning\cite{456} are also empirical ones. 
There is no doubt that empirical laws are very helpful for searching novel superconductors. 
However, the empirical laws are usually not so universal. 
There are often exceptions no matter the law was found by human or machine. 
Without physical picture, the empirical laws gave less contribution to theoretical studies.

Meanwhile, some other laws are known as physical parameter laws, which pursue physical meanings. 
Physical parameter laws indicate that the ${T}_\text{c}$ in diverse materials is correlated with some measurable macroscopic physical quantity, which is often further linked with some kind of carrier interaction, leading to a clue to the superconducting mechanism.

\begin{figure}%
\includegraphics{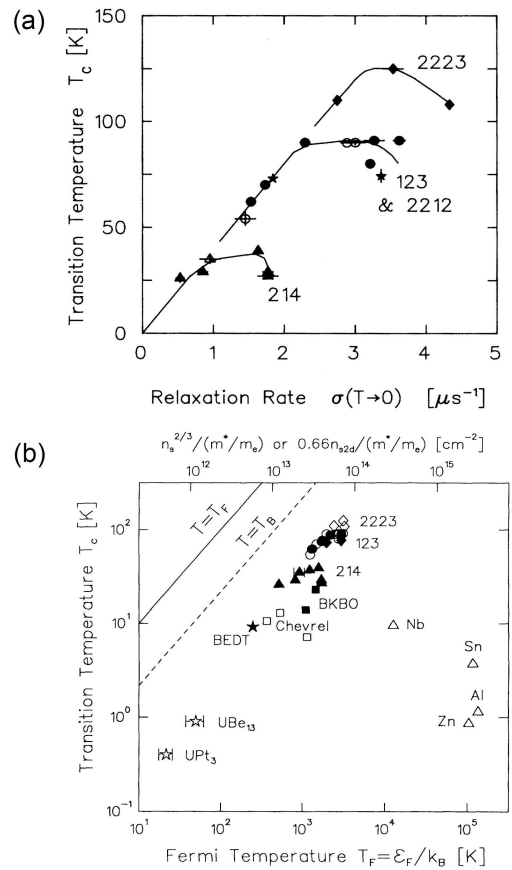}
\caption{
Uemura law. 
(a) ${T}_\text{c}$$\sim$$\sigma$ plot for cuprates as the doping content varies ($\sigma$$\propto$${n}_\text{s}$$/$${m}^\text{*}$)\cite{433}, 
(b)${T}_\text{c}$$\sim$${T}_\text{F}$ plot for diverse materials (${T}_\text{F}$$\propto$${n}_\text{s}$$/$${m}^\text{*}$ for 2D and ${T}_\text{F}$$\propto$${n}_\text{s}$$^{2/3}$$/$${m}^\text{*}$ for 3D)\cite{434}.
}
\label{FIG. Uemura law}
\end{figure}

Uemura law\cite{433, 434} is one of the well-known physical parameter laws. 
Soon after the discovery of HTS, Y. J. Uemura \textit{et al.} investigated the penetration depths ($\lambda$) of the cuprates by $\mu$SR measurements, and got a correlation of ${T}_\text{c}$$\sim$$1/$$\lambda$$^{2}$$\sim$${n}_\text{s}$$/$${m}^\text{*}$, where ${n}_\text{s}$ is the concentration of the superconducting carriers when $T$$\to$$0$ K, and the carrier effective mass (${m}^\text{*}$) is a parameter about the ability of the pairing interaction.

Uemura law has two angles of view. 
On the one hand, as shown in Fig.\ref{FIG. Uemura law}a, the ${T}_\text{c}$ varies with doping\cite{433}. 
On the other hand, as shown in Fig.\ref{FIG. Uemura law}b, the ${T}_\text{c}$ amongst diverse materials are different\cite{434}. 
When considering the influence of doping, all the under-doped cuprates are in agreement with a linear correlation of ${T}_\text{c}$$\propto$${n}_\text{s}$$/$${m}^\text{*}$, whereas the over-doped cuprates may show various behaviors. 
Deviation from the linear relationship appears to be very common for most over-doped cuprates, but there are also some cuprates obey the linear relationship even when they are heavily over-doped\cite{489, 490}. 
Meanwhile, when considering the material dependence of the ${T}_\text{c}$, cuprates, iron-based superconductors, heavy Fermion superconductors, and many unusual conventional superconductors, are in good agreement with a universal ${T}_\text{c}$$\sim$${T}_\text{F}$ correlation (${T}_\text{F}$ is the Fermi temperature), combining the cases of two and three dimension. 

Over the years, Uemura law has been supplemented and improved by a number of works\cite{439, 446}. 
More materials were added into the ${T}_\text{c}$$\sim$${T}_\text{F}$ correlation, even including the cases of the superfluid condensing of $^4$He and the Bose-Einstein condensing (BEC) of cold atoms\cite{439}. 
But, unfortunately, the ${T}_\text{c}$$\sim$${T}_\text{F}$ linear fitting does not work well for the simple substance superconductors. 

Uemura law gives an important clue to the nature of the superconductivity. 
Due to the superfluid density ($\rho_\text{s}$) is also proportion to ${n}_\text{s}$$/$${m}^\text{*}$ ($\rho_\text{s}$$=$$\mu_\text{0}$$e$$^2$${n}_\text{s}$$/$${m}^\text{*}$, based on London equations), Uemura law indicates the close link between ${T}_\text{c}$ and $\rho_\text{s}$.

\begin{figure}%
\includegraphics{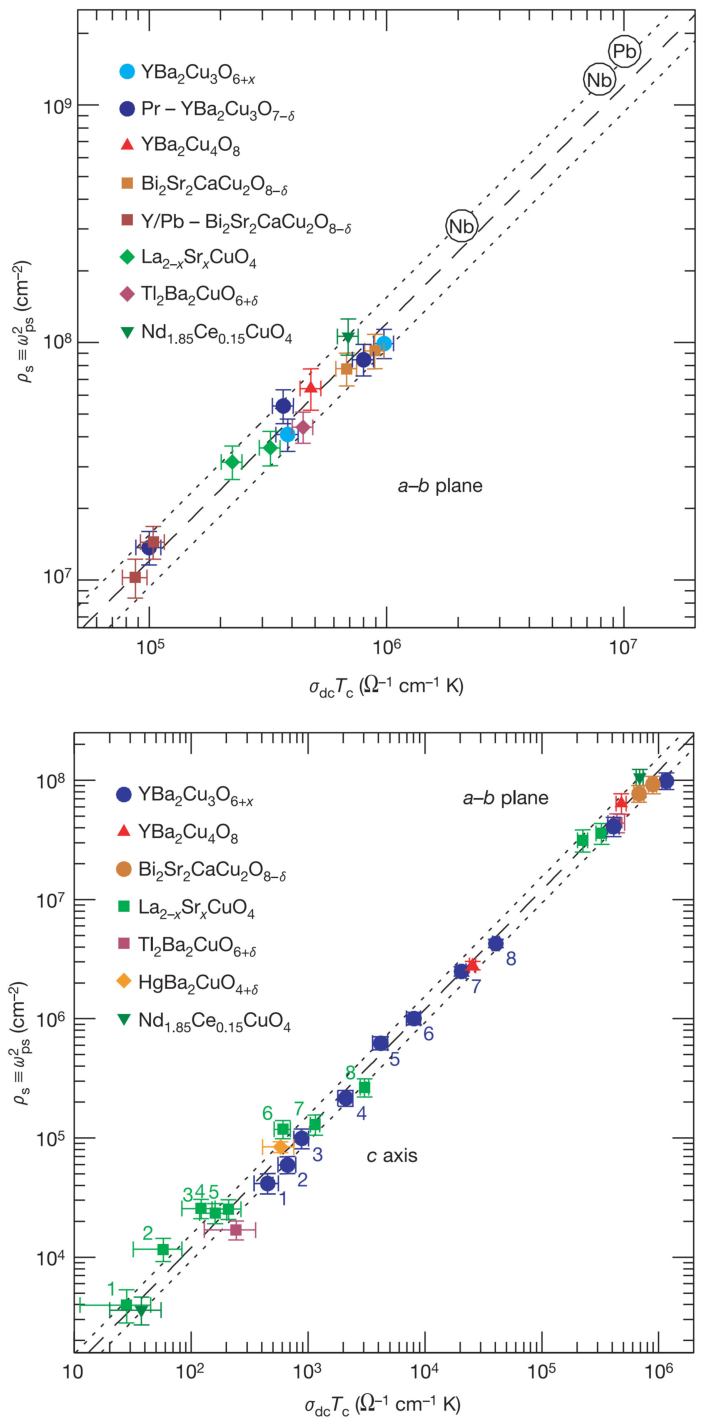}
\caption{Homes law\cite{394}.}
\label{FIG. Homes law}
\end{figure}

C. C. Homes \textit{et al.}\cite{394, 537} made an important progress after Uemura law. 
Homes law is a scaling relationship of $\rho_\text{s}$$\sim$$\sigma_\text{DC}$${T}_\text{c}$, where $\rho_\text{s}$ is the superfluid density when $T$$\to$$0$ K, and $\sigma_\text{DC}$ is the direct current conductivity slightly above the ${T}_\text{c}$ (see Fig.\ref{FIG. Homes law}).
Those over-doped cuprates, as mentioned above, which are not in good agreement with the linear trend in Uemura law, now can be well fitted by the $log($$\rho_\text{s}$$)$$\sim$$log($$\sigma_\text{DC}$${T}_\text{c}$$)$ straight line in Homes law. 
The universality of Homes law is good. 
It works well despite of the differences of crystalline structure, disorder, and anisotropy, $\textit{etc}$. 
However, deviation from the linear relationship still happens to some heavily over-doped cuprates\cite{455}, implying that some uncertain factor other than the amount of the superconducting carriers also matters. 
Homes law and Uemura law have a tight connection, thereby sometimes they are jointly called as ``Uemura-Homes law(s)''. 

As Uemura-Homes law(s) revealed the proportion relationship between $\rho_\text{s}$ and ${T}_\text{c}$, which is instructive for the understanding of the HTS, its connection with the BCS theory was discussed\cite{490, 516}. 
Further, J. Zaanen\cite{537} pointed out that Uemura-Homes law(s), as well as another poorly understood law, Tanner law\cite{552, 542}, are all related to a simple scaling behavior of Planckian dissipation, $\tau$$($${T}_\text{c}$$)$$\approx$$h/(2$$\pi$$k_\text{B}$${T}_\text{c}$$)$, which is a quantum physical constraint on the relaxation time ($\tau$). 
The high ${T}_\text{c}$, as well as the well-known $T$-linear resistivity, of the cuprates are attributed to it.

However, in Uemura-Homes law(s), $\rho_\text{s}$ and ${T}_\text{c}$ are both performance parameters of the superconductivity itself. 
The source of ${m}^\text{*}$ was not indicated. 
The information about the pairing interaction or glue is not explicit. 

Searching and identifying the pairing glue in high-${T}_\text{c}$ superconductors is an unfinished job. 
Owing that the condensing energy of the superconductivity is of a small order of magnitude ($\sim$0.01 eV), many kinds of electron interaction seems capable of providing a pairing glue strong enough.
Various elementary excitations and quasi-particles (\textit{e.g.} phonon, polaron, exciton, magnon, \textit{etc.}) have been considered. 
However, it turns out that not every kind of quasi-particle could act as the pairing glue. 
Moreover, P. W. Anderson \cite{38} demonstrated that there is no need of an additional boson quasi-particle to act as the pairing glue, because the Coulomb repulsion and the super-exchange interaction between the carriers are already adequate. 
By now, most candidates for pairing glue have been ruled out.
Yet the spin fluctuation is still promising. 
Usually, the spin fluctuation refers to the dynamics of the short-range magnetic order. 
It has been observed in cuprates, iron-based superconductors, heavy Fermion superconductors, and some other magnetic superconductors. 
In these years, the spin fluctuation in unconventional superconductors has received much attention\cite{472, 419, 15, 460, 50, 447}. 

\begin{figure*}%
\includegraphics{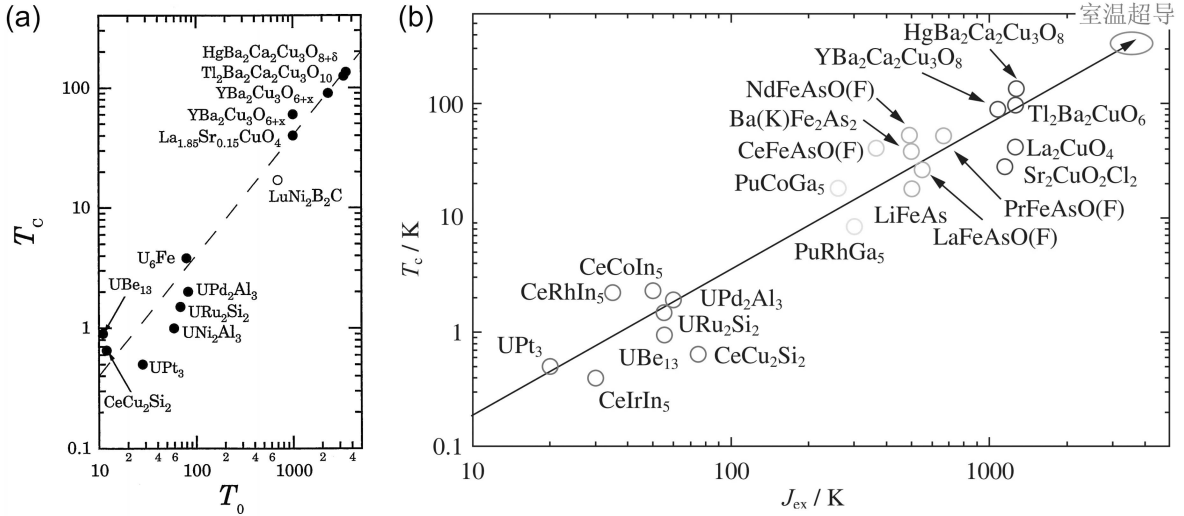}
\caption{
(a) The original version of Moriya-Ueda law\cite{436}, 
(b) An advanced version of ${T}_\text{c}$$\sim$${T}^\text{*}$ law\cite{448}.
}
\label{FIG. Moriya-Ueda law}
\end{figure*}

T. Moriya and K. Ueda \textit{et al.}\cite{436} developed a spin fluctuation model based on the self-consistent renormalization theory. 
Their model gave a parameter of characteristic temperature (${T}_\text{0}$), which expresses the frequency spread of the spin fluctuation. 
${T}_\text{0}$ is inversely proportional to the linear coefficient of the specific heat ($\gamma$), which can be estimated by the measurements of various macroscopic physical quantities\cite{466}. 
Then they obtained the dependence of the optimal ${T}_\text{c}$ on the ${T}_\text{0}$ (Fig.\ref{FIG. Moriya-Ueda law}a). 

Besides, for the heavy Fermion superconductors, there is another law (Fig.\ref{FIG. Moriya-Ueda law}b) showing the relevance between the ${T}_\text{c}$ and a characteristic temperature (${T}^\text{*}$) about the Kondo effect\cite{447, 448, 466, 449}. 
Y. F. Yang \textit{et al.}\cite{466} demonstrated that the essence behind the ${T}_\text{c}$$\sim$${T}^\text{*}$ law is the spin fluctuation and the antiferromagnetic exchange coupling. 
Recently, X. H. Chen \textit{et al.}\cite{449} found that the ${T}^\text{*}$ has a proportional relationship with the ${T}_\text{0}$ in Moriya-Ueda law, implying the same origin of them. 
In fact, ${T}_\text{0}$ is a parameter of the carrier behaviors, while ${T}^\text{*}$ is a parameter directly related to the pairing glue. 
They were both proposed to test the spin fluctuation models. 
Therefore, the ${T}_\text{c}$$\sim$${T}^\text{*}$ law and the Moriya-Ueda law could be categorized into one group. 
In the following, we call them ``Moriya-Ueda law(s)'' for convenience. 

Moriya-Ueda law(s) were also updated by the succeeding works. 
Soon after its discovery, the iron-based superconductors were added into the Moriya-Ueda law(s)\cite{448, 449}. 
In addition, the cuprates also can be added into a united law (Fig.\ref{FIG. Moriya-Ueda law}b) so long as converting the ${T}^\text{*}$ into the magnetic coupling strength (${J}_\text{ex}$)\cite{448}. 

Moriya-Ueda law(s) compared various materials at their optimal doping contents or pressure. 
However, in practical, the optimal doping contents or pressures of diverse materials are unequal, which is ignored in Moriya-Ueda law(s). 
Hence, strictly speaking, Moriya-Ueda law(s) are not rigorous quantitative scaling laws, but at most a trend or a sequence for different materials. 
Or to say, the Moriya-Ueda law(s) only worked well on the condition that the density of superconducting carriers in diverse materials were not very different. 
Even though, Moriya-Ueda law(s) were born for the spin fluctuation models. 
Its physical implement is explicit, which is an advantage that Uemura-Homes law(s) do not have. 
It is worth noting that the trend of the linear ${T}_\text{c}$$\sim$${T}_\text{F}$ relationship in Uemura law looks just like the trend in Moriya-Ueda law(s), although the physical meaning of them are quite different. 

In addition, there is a scaling relationship between the ${T}_\text{c}$ of the high-${T}_\text{c}$ superconductors and the resonance energy (${E}_\text{r}$) of magnetic excitation observed in neutron scattering experiments\cite{440, 441, 443, 452, 450}, which is also often seen as a proof of the spin fluctuation picture. 
The ${T}_\text{c}$$\sim$${E}_\text{r}$ law always works well, regardless of doping, crystal structure, type of disorder, and anisotropy, suggesting that ${E}_\text{r}$ could be regarded as an order parameter of phase transition for the HTS\cite{441}. 

By comparison, Uemura-Homes law(s) are derived from the superconducting state, Moriya-Ueda law(s) are derived from the normal state, and the ${T}_\text{c}$$\sim$${E}_\text{r}$ law is a phase transition critical behavior. 
According to ${T}_\text{c}$$\sim$${E}_\text{r}$ law and Moriya-Ueda law(s), the connection between the superconductivity and the spin fluctuation is obvious. 
However, the complete microscopic theory based on the spin fluctuation picture has not been established, yet. 
There are also options and possibilities other than the spin fluctuation picture\cite{60}. 
By and large, so far, the job of searching and identifying the pairing interaction in the HTS is far from over. 

Afterall, superconductivity is a unique physical phenomenon because it requires two necessary conditions: I. adequate itinerant carriers, II. each two carriers forming a pair by the aid of pairing interaction or glue. 
Correspondingly, it can be found from the laws mentioned above that the ${T}_\text{c}$ is governed by not one, but two factors, which are related to the amount of the carriers and the interaction between the carriers, respectively. 
And the two factors can be further decomposed into several parameters, including but not limited to the carrier concentration, the itinerant ability of the carriers, the intensity of the pairing glue, and the efficiency of the carrier-glue scattering. 
The right physical picture of the HTS could be obtained only by taking both aspects of factors into consideration, and knowing how each specific parameter affects the ${T}_\text{c}$. 

Unfortunately, in those ${T}_\text{c}$-involved laws already been found, either the material feature parameters with obvious physical implication are not easy to measure and calculate, or the material feature parameters easy to measure and calculate have no obvious physical implications. 
As a result, the criterion for the HTS mechanism is still in suspense. 

Actually, there has been a successful case for the conventional superconductors. 
The pairing mechanism based on electron-phonon coupling for the conventional superconductors was sustained by an evidence chain, consisting of the ${T}_\text{c}$ \textit{vs.} isotopic mass correlation, the phonon dispersion spectra, the electron-phonon coupling model, and the superconducting pairing correlation function spectra. 
Particularly, the isotopic mass experiments gave the information only about the pairing glue, because the carrier amount is unchanged rigidly. 
And, it is quite easy to associate the isotopic mass with the lattice vibration, which is how the electron-phonon coupling was identified as pairing glue. 
In fact, the isotopic mass is not a macroscopic physical quantity, but a structural parameter, as well as an intrinsic parameter of elements. 
So, the isotopic mass effect is neither a material empirical law, nor a physical parameter law, but a material gene law. 
It succeeded, not only because choosing the isotopic mass as a material gene parameter, but also owing to its clear indication of the pairing interaction. 
Nowadays, for the HTS, there is a lack of material gene laws like the isotopic mass effect, especially the laws with the band structure parameter. 

The significance of the band structure to the superconductivity is obvious. 
A law with band structure parameter would have a better chance to provide the key information about the microscopic mechanism of the HTS. 
Usually, it is expected that this band structure parameter should be found near the Fermi level. 
Accordingly, the Fermi surface morphology\cite{193, 67} and the orbital distribution near the Fermi level\cite{177, 4, 203} have been of great concern. 
However, for the high-${T}_\text{c}$ superconductors, the environment near the Fermi surface are so delicate that no universal ${T}_\text{c}$-involved law has been found, yet. 
New perspectives to search the material feature parameters in the band structure are required.

\section{Novel universal ${T}_\text{c}$ law with a band structure parameter}

\subsection{energy level distribution of electrons in diverse superconductors and its relevance to the ${T}_\text{c}$}

From the last section, it can be seen that in searching of universal laws, the key is to find an appropriate parameter with proper physical meaning. 
This parameter, should be associated with some kind of electron interaction, implying the identity of the pairing glue. 
And this parameter, better be derived from the band structural features, so as to reveal how the pairing glue works. 

Based on the concepts and ideas of big data and material genomics engineering, we surveyed hundreds of superconducting materials.
And owing that the energy scale of the superconducting transition is rather small, we don't want to overlook any kind of electron interaction, so as not to miss the vital one.
So, we are not going to focus just on the Fermi surface, but to begin with looking for the holistic features of the band structure in diverse superconductors, by comparing their energy level distribution of electrons.
All covalent electrons are considered, including the secondary-outer-shell electrons in saturated orbitals, as long as they participate in the orbital hybridization and coupling.

\begin{figure}%
\includegraphics{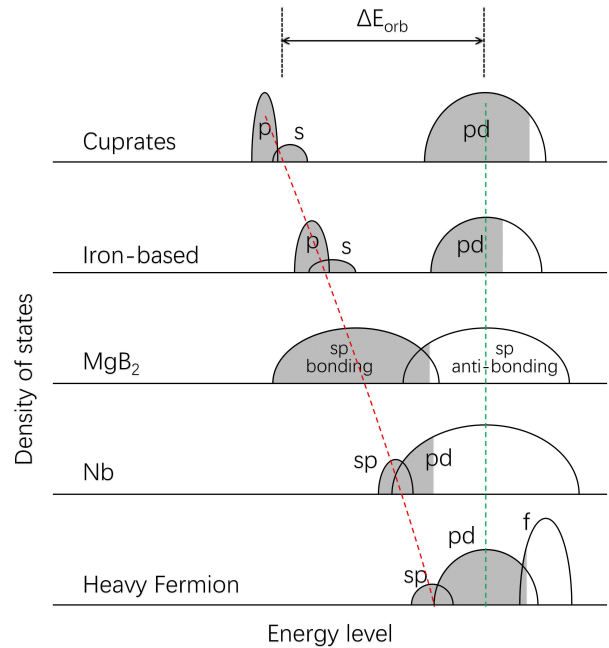}
\caption{
Holistic characteristics of the band structure of various superconductors (schematic illustration of the density of states graphs). 
Shadow areas represent occupied orbitals. 
The red dash line is a guide to the eye showing the energy level variance of the hybridized secondary outer orbitals.
}
\label{FIG. BandFeature}
\end{figure}

Figure \ref{FIG. BandFeature} shows the holistic characteristics of the band structure of cuprates (\textit{e.g.} \text{Sr$_{1-x}$Ca$_{x}$CuO$_{2}$}), iron-based superconductors (\textit{e.g.} \text{BaKFe$_{4}$As$_{4}$}), conventional superconducting compounds (\textit{e.g.} \text{MgB$_{2}$}), conventional superconducting elements (\textit{e.g.} Nb), and heavy Fermion superconductors (\textit{e.g.} \text{UPt$_{3}$}), in order of the ${T}_\text{c}$. 
It can be seen that there is a group of \textit{s-p} hybridized secondary outer orbitals, moving towards deeper energy level with the increment of the ${T}_\text{c}$. 
The trend is clear and easy to be verified.
Then we define a parameter, $\Delta$${E}_\text{orb}$, as the energy level spacing between the outermost orbitals and the secondary outer orbitals, and compare hundreds of superconducting materials by their $\Delta$${E}_\text{orb}$ and ${T}_\text{c}$. 

\begin{figure*}%
\includegraphics{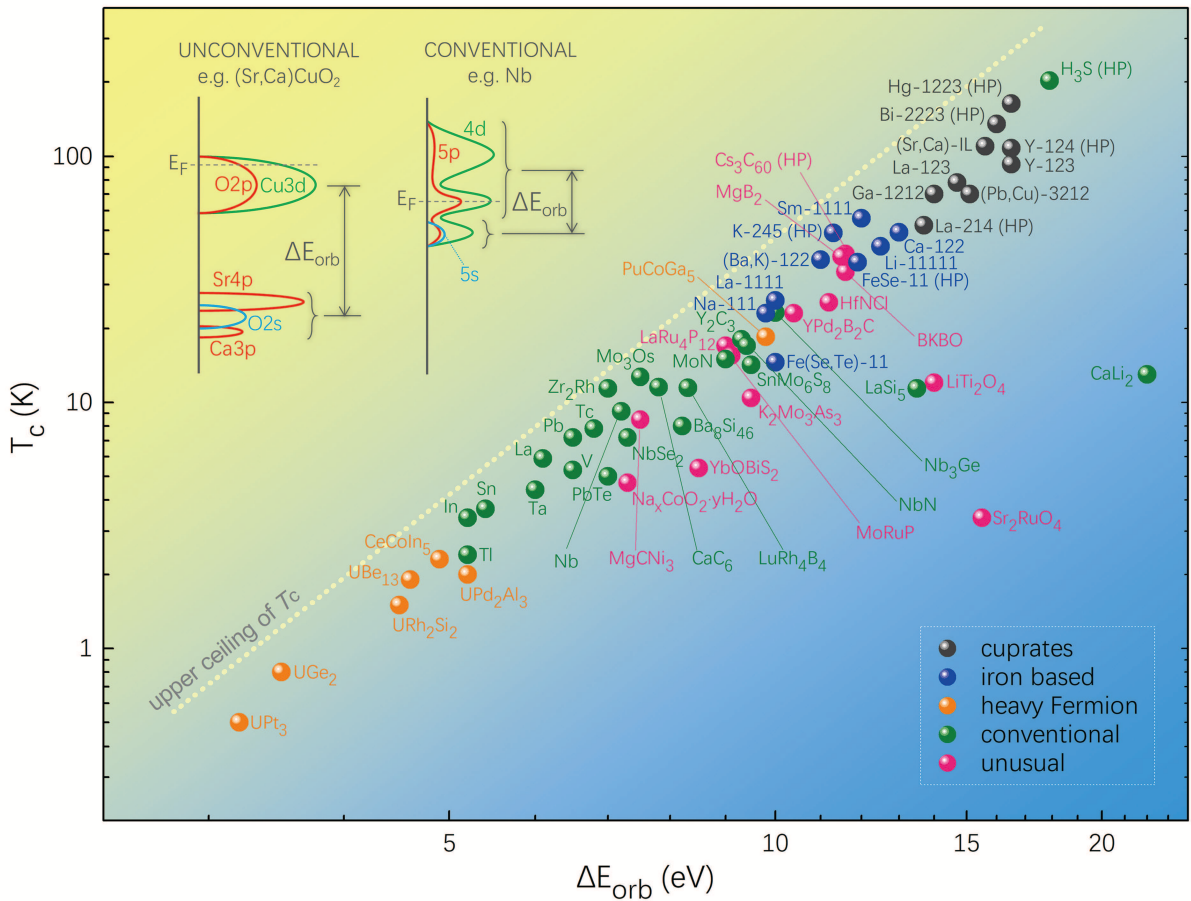}
\caption{
Dependence of the ${T}_\text{c}$ on the band structure parameter, $\Delta$${E}_\text{orb}$ (in log-log coordinates). 
$\Delta$${E}_\text{orb}$ is the energy level interval between the outermost orbitals and the hybridized secondary outer orbitals, which is derived from first principle calculation. 
The top-left insets are the schematic density of states graphs of two typical materials, which shows how the $\Delta$${E}_\text{orb}$ is derived. 
Here, the ${T}_\text{c}$ values of all unconventional superconductors are the optimal ${T}_\text{c}$. 
The dash line is a guide to the eye, implying a ceiling of the ${T}_\text{c}$ for diverse superconductors.
}
\label{FIG. Our Law}
\end{figure*}

Figure \ref{FIG. Our Law} shows the relevance between the optimal ${T}_\text{c}$ and the $\Delta$${E}_\text{orb}$ for different kinds of superconductors. 
The value of the optimal ${T}_\text{c}$ were mostly obtained from the NIMS superconducting database ({http://supercon.nims.go.jp/}). 
The $\Delta$${E}_\text{orb}$ data were derived from the calculation using CASTEP\cite{499}. 
The energy levels of the outermost and the secondary outer orbitals were estimated from the partial density of states (PDOS) results. 
It can be seen in Fig.\ref{FIG. Our Law} that the ${T}_\text{c}$ varies for over two orders of magnitude when the $\Delta$${E}_\text{orb}$ varies for nearly ten times. 
The light color dash line in Fig.\ref{FIG. Our Law} qualitatively shows a trend: the ${T}_\text{c}$ of each kind of superconductors has an upper limit, which at large has a power function relationship with the $\Delta$${E}_\text{orb}$. 
Most data points in Fig.\ref{FIG. Our Law} appear close to the dash line (${T}_\text{c}$ upper limit), because they belong to the superior superconductors, where ``superior'' means that these superconductors have relatively high ${T}_\text{c}$ in their own material families. 
Meanwhile, much more superconductors with lower ${T}_\text{c}$ would appear beneath the dash line (not shown in the figure for clarity). 

Please note that the so-called ``outermost/secondary-outer'' here means high/low in energy level, instead of outside/inside in coordinate space. 
More specifically, the outermost orbitals refer to a group of lowest unsaturated (not fully occupied) orbitals, such as Cu\textit{3d} and O\textit{2p} in cuprates. 
While the secondary outer orbitals refer to a group of highest saturated (fully occupied) orbitals, which usually form \textit{s-p} hybridization. 
For instance, as shown in the inset of Fig.\ref{FIG. Our Law}, for the \text{(Sr,Ca)CuO$_{2}$} (an infinite layer cuprate), the secondary outer orbitals consist of O\textit{2s} and Sr\textit{4p}(Ca\textit{3p}). 
And, for the Nb simple substance, \textit{4d}, \textit{5s} and \textit{5p} orbitals together form the conduction band, in which there is an \textit{s-p} hybridization part with lower energy level and a \textit{p-d} hybridization part with higher energy level. 
We consider the \textit{s-p} hybridization part as the secondary outer orbitals, and the \textit{p-d} hybridization part as the outermost orbitals, respectively. 
In addition, for the covalent compounds \text{MgB$_{2}$} (see Fig.\ref{FIG. BandFeature}, row 3), which also needs to consider the orbital hybridization and coupling first, the bonding state are of the secondary outer orbitals, and the antibonding state are of the outermost orbitals. 

\begin{figure}%
\includegraphics{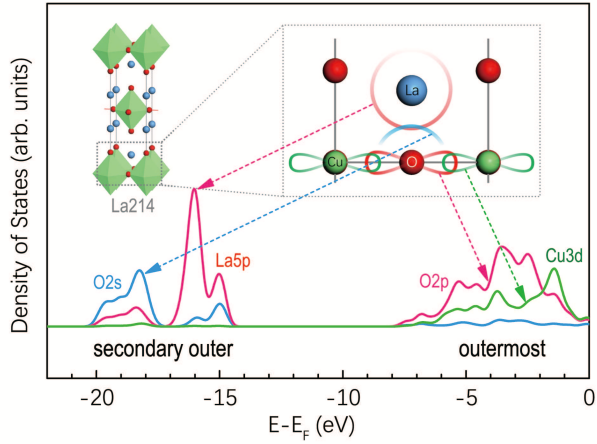}
\caption{
The orbital distribution in a typical cuprate (La-214). 
The outermost orbitals consist of Cu\textit{3d} and O\textit{2p}, while the secondary outer orbitals consist of O\textit{2s} and La\textit{5p}.
}
\label{FIG. DOS and orbital interaction}
\end{figure}

When defining the outermost and the secondary outer orbitals, both the energy level and the hybridization are considered. 
The outermost and the secondary outer orbitals often participate in different types of bonding and interaction. 
In simple substances, the outermost orbitals form metallic bond, while the secondary outer orbitals form covalent bond. 
Whereas, in cuprates, such as La-214 (see Fig.\ref{FIG. DOS and orbital interaction}), the outermost orbitals (O\textit{2p} and Cu\textit{3d}) form covalent bonding and generate the super-exchange interaction, while the secondary outer orbitals (O\textit{2s} and La\textit{5p}) make considerable contribution to the van der Waals (VDW) interaction. 

\subsection{physical implications of the band structure parameter $\Delta$${E}_\text{orb}$ and the ${T}_\text{c}$$\sim$$\Delta$${E}_\text{orb}$ correlation}

Our ${T}_\text{c}$$\sim$$\Delta$${E}_\text{orb}$ correlation gives the information about the pairing interaction rather than about the carrier concentration and density, which is similar to the cases of Moriya-Ueda law(s). 
So, we should look deeper into the electronic interaction represented by the $\Delta$${E}_\text{orb}$. 
Unlike Moriya-Ueda law(s) indicating magnetic coupling and spin fluctuation, our ${T}_\text{c}$$\sim$$\Delta$${E}_\text{orb}$ correlation indicates VDW interaction and fluctuation of orbital hybridization. 

The VDW interaction used to be ignored by default in the research works about superconductivity. 
However, we demonstrate that the influence of the VDW interaction in high-${T}_\text{c}$ superconductors was underestimated. 
It should be emphasized that the VDW interaction exists everywhere, and its energy scale ($\sim$0.1 eV) is no less than the condensing energy of the HTS. 
Such an important interaction has been long overlooked inappropriately. 

It is well accepted that the VDW interaction between adjacent atoms derives from the coupling of atomic transient dipole moments. 
Usually, when discussing the VDW interaction, each atom is looked as a whole. 
But if we go deeper, it will be found that the atomic transient dipole moment is a result of collective effect, derived from the excitation of electrons in each orbital, by means of the change of atomic orbital hybridization (AOH) between the occupied orbitals and the empty orbitals. 

It may be not well noticed that for the electrons in various orbitals, the ability to be excited are very different, resulting in unequal contribution to the atomic transient dipole moment. 
The core electrons at very deep energy levels are restricted by the nucleus, while the covalent electrons at very shallow energy levels are restricted by the chemical bond (\textit{i.e.} exchange coupling). 
Their excitations are suppressed. 
So, the contributions of them to the AOH changing are suppressed.
Then, only the electrons in moderately deep energy levels are capable of having dramatic change of AOH. 
Consequently, amongst the orbital electrons in different energy levels, the electrons in the secondary outer orbitals make the major contribution to the transient dipole moments as well as the VDW interaction. 
The magnitude of transient dipole moments and VDW interaction could be roughly estimated by seeing the secondary outer orbitals. 

For the electrons in secondary outer orbitals, whose energy levels ($E$$-$${E}_\text{F}$) range from 0 eV to $-$20 eV, the restriction on their AOH change comes mainly from the chemical bonds. 
The deeper the energy levels, the weaker the restriction, the more intense the AOH change, the larger the transient dipole moment, and the stronger the VDW interaction. 

Further, owing that the secondary outer orbitals are usually saturated, the secondary outer electrons cannot fluctuate alone. 
But they can fluctuate with the outermost electrons by altering the ratio of orbital hybridization. 
For example,  in hole-doped La-214 cuprates, given the ground state of the O anion is $|$$\textit{2s}^{2}$$>$$+$$|$$\textit{2p}^{5.9}$$>$ and the transient excited states are $|$$\textit{2s}^{2-\delta}$$\textit{2p}^{\delta}$$>$$+$$|$$\textit{2s}^{\delta}$$\textit{2p}^{5.9-\delta}$$>$, the motion of the transient dipole moment can be comprehended as a vibration of $|$$\textit{2s}^{2}$$>$$+$$|$$\textit{2p}^{5.9}$$>$ $\longleftrightarrow$ $|$$\textit{2s}^{2-\delta}$$\textit{2p}^{\delta}$$>$$+$$|$$\textit{2s}^{\delta}$$\textit{2p}^{5.9-\delta}$$>$ by changing the AOH ratio ($\delta$).
Likewise, the La cation could have a vibration of $|$$\textit{5p}^{6}$$>$$+$$|$$\textit{6s}^{0}$$>$ $\longleftrightarrow$ $|$$\textit{5p}^{6-\delta}$$>$$+$$|$$\textit{6s}^{\delta}$$>$.
Thus, in high-${T}_\text{c}$ superconductors, although the saturated secondary outer orbitals cannot provide carriers, but \textit{they do can influence the carriers by hybridization fluctuation} (not the fluctuation of electron hopping in the energy band, but the fluctuation of the hybridization ratio in each atom). 
The collective fluctuation of AOH in the whole crystal leads to an indirect interaction of the carriers and a long-range correlation through the VDW interaction, which is strong enough to support the pairing in HTS. 

Thus, the ${T}_\text{c}$$\sim$$\Delta$${E}_\text{orb}$ positive correlation shown in Fig.\ref{FIG. Our Law} could be understood as:
Larger $\Delta$${E}_\text{orb}$ leads to larger transient dipole moments and stronger VDW interaction, as well as higher intensity of the collective hybridization fluctuation, then stronger pairing interaction, then higher ${T}_\text{c}$. 

\subsection{competences of the ${T}_\text{c}$$\sim$$\Delta$${E}_\text{orb}$ correlation}

Firstly, our ${T}_\text{c}$$\sim$$\Delta$${E}_\text{orb}$ correlation is an analogue of isotopic mass effect. 
The $\Delta$${E}_\text{orb}$ is a parameter not only related to carrier interaction, but also related to the band structure. 
The band structure information offers a better chance to identify the pairing glue, and eventually lead to the microscopic mechanism of the HTS. 

Secondly, the universality of our ${T}_\text{c}$$\sim$$\Delta$${E}_\text{orb}$ correlation is good. 
The material-dependent trends presented in Fig.\ref{FIG. Our Law} is a statistic result, which covers all kinds of superconductors.
Besides, the parameter $\Delta$${E}_\text{orb}$ is derived from the distribution of orbital energy levels in each atom, which is one of the intrinsic properties of elements. 
For a certain material, the value of the $\Delta$${E}_\text{orb}$ only depends on the chemical composition (element type), the ionic valence, and the atomic coordinate. 
Whereas, it does not rely on doping, Fermi surface, pairing symmetry, methods of measurement or calculation, \textit{etc.} 
Therefore, our ${T}_\text{c}$$\sim$$\Delta$${E}_\text{orb}$ correlation is actually an element-dependent law, which is literally universal. 

Thirdly, we can use the ${T}_\text{c}$$\sim$$\Delta$${E}_\text{orb}$ correlation to predict the ${T}_\text{c}$ limit of new superconductors. 
The value of the $\Delta$${E}_\text{orb}$ can be obtained either experimentally or theoretically. 
Any material, known as superconductor or not, can be easily verified by first principle calculations. 
Since it does not care about the details near the Fermi surface, there is no need to worry about the strong-correlation problem, even the approximation methods based on density functional theory (DFT) can meet the demand of estimating the value of $\Delta$${E}_\text{orb}$. 
As a statistic result, the ${T}_\text{c}$$\sim$$\Delta$${E}_\text{orb}$ correlation does not require high accuracy of each single datum. 
We must admit that in the present stage the ${T}_\text{c}$$\sim$$\Delta$${E}_\text{orb}$ correlation is not well quantified. 
The parameter $\Delta$${E}_\text{orb}$ is just a rough estimation of the strength of VDW interaction. 
There were human-induced uncertainties in defining and reading the value of the $\Delta$${E}_\text{orb}$. 
The calculation processes were inevitably subject to approximations and errors. 
However, despite of all those imperfections, the general trend of the ${T}_\text{c}$$\sim$$\Delta$${E}_\text{orb}$ correlation is robust and liable. 

In addition, it should be emphasized that both in a universal law like our ${T}_\text{c}$$\sim$$\Delta$${E}_\text{orb}$ correlation, it is unappropriated to use the linear fitting line, because the physical quantity of the horizontal-axis in these laws is not the only parameter governing the ${T}_\text{c}$. 
As mentioned above, the ${T}_\text{c}$ is affected by multiple factors (including the amount of the carriers, which is not considered in our ${T}_\text{c}$$\sim$$\Delta$${E}_\text{orb}$ correlation). 
For example, the ${T}_\text{c}$ value of a cuprate would reach its maximum only when the doping-related factors are optimized, or else, the ${T}_\text{c}$ would be below the optimal value. 
Meanwhile, for other superconducting materials, their ${T}_\text{c}$ might not reach the maxima predicted by Fig.\ref{FIG. Our Law}, for a variety of reasons. 
Therefore, when comparing diverse superconducting materials, we can only see a rough upper ceiling line of the optimal ${T}_\text{c}$. 
In a correlation plot like Fig.\ref{FIG. Our Law}, the data points reaching the ceiling are actually the minority, while a lot of data points should spread in the broad range below the ceiling line. 
Then, the ceiling line would not appear clearly until collecting sufficient data of the superior materials with higher ${T}_\text{c}$. 
The position of the ceiling could be known more precisely from big data analyses dealing with all known superconductors. 

Also, a parameter better than $\Delta$${E}_\text{orb}$ is desired to estimate the intensity of the hybridization fluctuation and the strength of VDW interaction more precisely. 
Further, direct detection of the hybridization fluctuation in experiment are highly recommended.

\section{Summary}

The ${T}_\text{c}$-involved laws, especially the relevance between the ${T}_\text{c}$ and band structure parameters are of great significance to explore the HTS mechanism or new superconductors.
We report a universal correlation between the ${T}_\text{c}$ and the energy level of electrons. 
It implies the fluctuation of orbital hybridization is a new candidate of pairing glue, and the van der Waals interaction may play a vital role in the high-${T}_\text{c}$ superconductivity. 

\begin{acknowledgments}
This work is supported by the Ministry of Science and Technology Key Special Project of China, ``Engineering Special Database and Material Big Data Technology for Material Genome'': 2016YFB0700503-7.
\end{acknowledgments}

\bibliographystyle{unsrturl}

\bibliography{ref}

\end{document}